\documentstyle[12pt]{article}
\input psfig
\input rotate
 1
\def\as{\alpha_{\rm S}}

\def\citenum#1{{\def\@cite##1##2{##1}\cite{#1}}}
\def\citea#1{\@cite{#1}{}}

\def\as{\alpha_{\rm S}}

\def\to{\rightarrow}

\def\a{\alpha}

\def\D{\Delta}

\def\g{\gamma}
\def\G{\Gamma}

\def\l{\lambda}

\def\o{\omega}

\def\pa{\partial}

\def\ra{\rightarrow}

\def\s{\sigma}

\def\({\left(}
\def\){\right)}

\def\citenum#1{{\def\@cite##1##2{##1}\cite{#1}}}
\def\citea#1{\@cite{#1}{}}

\def\l1vt{\vec{l_{1\perp}}}

\def\rt{r_{\perp}}
\def\bt{b_{\perp}}
\def\rt2{$r^2_{\perp}$}
\def\bt2{$b^2_t$}
\def\aa{$z(1 - z)\, Q^2\,\,+\,\,m^2_Q$}

\def\jol1{$J_0(\,l_{1\perp}\,r_{\perp}\,)$}
\def\ko{K_0(\,a\,r_{\perp}\,)}

\def\citea#1{\@cite{#1}{}}

\def\VEV#1{\left\langle #1\right\rangle}

\def\beq{\begin{equation}}
\def\eeq{\end{equation}}
\def\bea{\begin{eqnarray}}
\def\eea{\end{eqnarray}}

\def\bbbz{{\mathchoice {\hbox{$\sf\textstyle Z\kern-0.4em Z$}}
{\hbox{$\sf\textstyle Z\kern-0.4em Z$}}
{\hbox{$\sf\scriptstyle Z\kern-0.3em Z$}}
{\hbox{$\sf\scriptscriptstyle Z\kern-0.2em Z$}}}}
\def\npb#1#2#3{    {\it Nucl. Phys. }{\bf B#1} (19#2) #3}
\def\plb#1#2#3{    {\it Phys. Lett. }{\bf B#1} (19#2) #3}
\def\prd#1#2#3{    {\it Phys. Rev. }{\bf D#1} (19#2) #3}

\def\prl#1#2#3{    {\it Phys. Rev. Lett. }{\bf #1} (19#2) #3}

\def\zpc#1#2#3{    {\it Z. Phys. }{\bf C#1} (19#2) #3}

\def\sjnp#1#2#3{   {\it Sov. J. Nucl. Phys. }{\bf #1} (19#2) #3}

\relax
\begin{document}
\begin{titlepage}
\noindent
 August  1995   \hfill  CBPF-NF-053/95 \,\,\,\, TAUP\,\,2283/95\\[3ex]
\begin{center}
{\Large\bf SHADOWING CORRECTIONS     }   \\[1.4ex]
{\Large \bf IN DIFFRACTIVE  QCD LEPTOPRODUCTION  }\\[1.4ex]
{\Large\bf OF VECTOR MESONS }  \\[7ex]
{\large E.\ Gotsman} $^{a)}$\
\footnote {e-mail: gotsman@ccsg.tau.ac.il},\
{\large E.M.\ Levin} $^{b)}{}^{\ c)}$\
\footnote {e-mail: levin@lafex.cbpf.br}\
{\large and\ U.\ Maor} $^{a)}{}^{\ b)}$
\footnote {e-mail: maor@vm.tau.ac.il} \\[4.5ex]
a) { School of Physics and Astronomy  }      \\
{Raymond and Beverly Sackler Faculty of Exact Sciences} \\
{Tel Aviv University, Tel Aviv 69978, ISRAEL}  \\ [1.5ex]
b) { LAFEX, Centro Brasileiro de Pesquisas F\'\i sicas  (CNPq)}\\
{Rua Dr. Xavier Sigaud 150, 22290 - 180 Rio de Janeiro, RJ, BRASIL}\\[1.5ex]
c) { Theory Department, Petersburg Nuclear Physics Institute}\\
  { 188350, Gatchina, St. Petersburg, RUSSIA}\\[7.5ex]
\end{center}
{\large \bf Abstract:}
The formulae for shadowing corrections in deep inelastic
leptoproduction of vector mesons are presented. These formulae are
also applicable for photoproduction of vector
mesons constituted of heavy quarks.
Our results are conveniently presented by
the definition of a damping factor
giving  the reduction of the calculated cross sections due to
shadowing. Our calculated cross sections are compared with those
obtained with no shadowing and with the available experimental data
including the recent data from HERA. We have also investigated
the importance of shadowing on
the relationship between
  $\frac{\pa F_2(Q^2,x)}{ \pa ln Q^2}$ and the cross section for
 virtual photoproduction of vector mesons.
A discussion of shadowing corrections
to the proton's gluon density is presented and  numerical
estimates are given.

\end{titlepage}

\section{ Introduction }
Over the last decade, following the paper of
 Bartels and Loewe[1], diffractive leptoproduction of vector
mesons has been investigated within the framework of perturbative
QCD (pQCD). In particular, it has been shown[1,2]
that the DIS cross sections can be calculated in
pQCD provided that both the energy and photon virtuality are large enough.
These cross sections are proportional to
$(x G(Q^2,x))^2$, where $x G(Q^2,x)$ is the gluon distribution
within the nucleon target. Futhermore, Donnachie and Landshoff[3]
have shown in a non perturbative Pomeron approach, that
both the initial photon and the produced vector meson have to be
longitudinally polarized  to realize
the leading $Q^2$ behaviour of the cross section
 i.e. $\frac{d \s}{d t} \propto \frac{1}{Q^4}$. This result has also
 been confirmed in the BFKL approach to the Pomeron structure in pQCD[4].
A new insight into the problem has been suggested by Ryskin[5], who
proved that J/$\psi$ leptoproduction can be   calculated reliably
in the leading log approximation (LLA)
of pQCD, and that the non-perturbative effects coming from
 large distances, can be factored out in terms of the wave function
of the produced vector meson at the origin.
In addition, Ryskin  pointed out that the pQCD calculation is also
 valid for leptoproduction of light vector mesons. For heavy
vector mesons the calculation can be safely continued to $Q^{2}$ = 0.
  An important contribution in  the construction of this
formalism appears in the paper of
 Kopeliovich et.al.[6], where the wave function for the light
vector mesons is evaluated using the constituent quark model.
Finally, we mention the contribution of
 Brodsky et.al.[7] who showed how to factor out the long
distance effects for the case of light vector mesons in terms of the light-cone
$\bar q q $ wave function of the produced vector meson[8].

The goal of this paper is to study the shadowing (screening)
corrections (SC)
for  lepto and photoproduction of vector meson. To this end we shall
generalize the formalism developed by Mueller[9]
for the gluon and quark densities. Technically, we shall utilize
the idea of a $r_{\perp}$- representation for the SC
 suggested in Ref.[10] and further discussed
in Refs.[6,9,11,12]. The quantity we wish to calculate is the
damping factor defined in our previous publication[13] as:
\beq \label{DF}
D^2\,\,=\,\,\frac{\frac{d \s(\g^* p \ra   VP)}
{d t} [\,\, with \,\,\,SC]}
{\frac{d \s( \g^* p \ra  Vp)}{d t}[ \,\,without\,\,\,SC]}\,\mid_{t = 0}
\eeq
where $ t = - q^2_{\perp}$ denotes  the square of
 momentum transfer in the reaction
$ \gamma^{*} p \rightarrow V p $ .

\par The structure of our paper is as follows:
In section 2 we define our notation and numerical coefficients
pertinent to the DIS production of vector mesons without including SC.
 Basically, we reproduce the well known results of Refs.[5,7],
doing calculations in the $r_{\perp}$-representation ( $r_{\perp}$
denotes the transverse separation between
the quark and antiquark).

 We will
show that the results hold, not only in the leading log
 of the GLAP evolution equations [14], but also in the
leading ln$\frac{1}{x}$ approximation of pQCD, as was pointed out by
Ryskin[5].
In section 3 we extend the  formalism  suggested
 by Levin and Ryskin[10] and Mueller[9] to the case of
  a $\bar q q $ pair with a definite value of $r_{\perp}$ transversing
the target (be it a proton or nucleus). We will show that for
 vector meson production we obtain a closed expression for the
damping factor. Our result does not depend on the large  transverse
distances, allowing us to calculate the SC to a good theoretical
  accuracy in pQCD.
In the fourth section we prove that the damping factor can be rewritten
using the experimentally
  measured value of $\frac{ \pa F_2(Q^2,x)}{ \pa \ln Q^2}$. To this end,
we use the leading log $\frac{1}{x}$ approximation of pQCD with
a variety of shadowing corrections.
In  section 5 we  discuss the SC for the
gluon distribution, and investigate the uncertainties due to
 large distance contributions which can be important
for  this case.
Section 6 contains the numerical evaluations of our formalism, which
are compared with those obtained for the case with no SC.
Our results are then compared with
the experimental data,
mostly from HERA.
A summary  and final discussion are presented in section 7.

\section{ Vector meson production in DIS without shadowing corrections}
\subsection{Notations}
In this paper we will use the  following notations (see Fig.1):

  $Q^2$ denotes the virtuality of the photon in DIS, $m_V$ is
the mass of the produced vector meson and m is the target mass assumed to be
a proton unless specified otherwise.

 $x \,\,=\,\,\frac{Q^2 + m^2_V}{s}$,
where $\sqrt{s}$ is the c.m. energy of the
incoming photon-proton system.

 $ a^2 $= \aa, where $z$  is the fraction of of the photon energy
carried by the quark. $m_Q$ is the mass of the current quark.

$\vec{k}_{\perp}$ denotes the transverse momentum of the quark, and
$\vec{r}_{\perp}$  the transverse
separation between the quark and antiquark.

$\vec{b}_{\perp}$ is the impact parameter
 of our reaction which is the variable conjugate
to the momentum transfer ($\vec{q}_{\perp}$).

 $\vec{l}_{i\perp}$ denote the transverse momentum of the gluons
 attached to the quark-antiquark pair (see Fig.1).

 We will use the evolution equations for the parton densities
in the moment  space. For any function $f(x)$, we
define a moment $f(\o)$
\beq \label{MOM}
f (\o)\,\,=\,\,\int^1_0\,dx \,\,x^{\o}\,\,f(x)\,\,\,\,.
\eeq
Note that,
the moment variable $\o$, is chosen such that the  $\o$ = 0 moment
measures the number of partons, and the $\o$ = 1 moment measures their
momentum. An alternative moment variable N defined such that N = $\o$ + 1
, is often found in the literature.

The $x$ - distribution can be reconstructed by considering the inverse
Mellin transform. For example, for the gluon distribution it reads
\beq \label{IMEL}
x G(Q^2,x)\,\,=\,\,\frac{1}{2 \pi i}\,\,\int_C \, d \,\o \,\,x^{- \o}
\,\,g (Q^2,\o)\,\,.
\eeq
The contour of integration C is taken to the right of all singularities.
The solutions to both the GLAP and the BFKL equations have the general form
\beq
g(Q^2,\o)\,\,=\,\,g(\o)\,e^{\g(\o) ln Q^2}\,\,,
\eeq
where $\g(\o)$ denotes the anomalous dimension, which  in the leading
 ln$\frac{1}{x}$ approximation of pQCD, is
a function of $\frac{\as}{\o}$ and can be
 presented as the following series[15]
\beq \label{LIANDI}
\g ( \omega ) \,\,=\,\,\frac{\as N_c }{\pi} \frac{1}{\omega} \,\,+\,\,
\frac{2  \as^4 N_c^4 \zeta ( 3 )}{\pi^4} \frac{1}{\omega^4}\,\,+\,\,O
( \frac{\as^5}{ \omega^5} )\,\,.
\eeq

 Our amplitude is normalized such that
$$
\frac{d \s}{d t}\,\,=\,\,\pi \,|f(s,t)|^2 \,\,,
$$
$$
\s_{tot}\,\,=\,\,4\pi\,\,Im\,f (s,0)\,\,.
$$
The scattering amplitude in $b_{\perp}$-space is defined as
$$
a(s,b_{\perp})\,\,=\,\,\frac{1}{2 \pi}\,\,\int\,d^2 q_{\perp} \,\,
e^{- i {\vec{q}}_{\perp}\,\cdot\,{\vec{b}}_{\perp}}\,\,f(s,t = - q^2_{\perp})
\,\,.
$$
In this representation
$$\s_{tot}\,\,=\,\,2\,\,\int \,d^2 b_{\perp} \,\,Im\,\,a (s, b_{\perp} )\,\,,
$$
$$
\s_{el}\,\,=\,\,\int \,d^2 b_{\perp} \,\,|\,a (s, b_{\perp} )|^2\,\,.
$$

In general we  have attempted to adjust our notation and normalization
to agree with those of
 Brodsky et.al.[7].

\subsection{The amplitude in the $r_{\perp}$ representation}
This approach was originally formulated in Ref.[10] and
has been carefully developed in Ref.[9].
While the boson projectile is transversing
  the target, the distance $r_{\perp}$ between the quark and
antiquark can vary by an amount
  $\D r_{\perp}\,\,\propto R \,\,\frac{k_{\perp}}{E}$,
where $E$ denotes the energy of the pair in the target rest frame,
and  R  is the size
of the target (see Fig.1).
 The quark
transverse momentum is  $k_{\perp} \,\propto \,\frac{1}{r_{\perp}}$. Therefore
\beq
\D r_{\perp}\,\,\propto\,\,R \,\frac{k_{\perp}}{E}\,\,\ll\,\,r_{\perp}\,\,,
\eeq
which is valid if
\beq
r^2_{\perp} \cdot s \,\gg 2 m \,R\,\,,
\eeq
where $s \simeq \ 2 m E $.
The above condition has the following form in terms of $x$
\beq
x\,\,\ll\,\,\frac{1}{ 2 m R }\,\,.
\eeq
Hence, at small values of $x$, the transverse distance
between the quark and antiquark
 is a good degree of freedom[9,10,16], and
the interaction of a virtual photon with the target can be written in the form
( we use the notation of Ref.[7])
\beq
M_f\,\,=\,\,\sqrt{N_f}\,\,\sum_{\lambda_1 \,\lambda_2}
 \int^1_0 \,d\,z\,\,\int\,\frac{d^2 r_{\perp}}{2 \pi}\,\,\,
\Psi^{\g^*}(Q^2,r_{\perp},z)
\,\,\s(r_{\perp},q^2_{\perp})\,\,
{[{\Psi^V}(r_{\perp},z)]}^*\,\,,
\eeq
where $\lambda_i$ denotes the polarization of the quarks.
Note that both $\g^*$ and $V$ are longitudinally polarized.
$\Psi^V$ denotes the wave function of the produced vector meson,
and we anticipate that the value of
 $r_{\perp}$ that dominates the integral in Eq.(9) will be small
(of the order of $r_{\perp}\,\propto\,\frac{1}{Q}$).
The form of $\Psi^{\g^*}$, the wave function of the
longitudinally polarized photon,
 has been given in Ref.[17], i.e.
\beq
\Psi^{\g^*}(r_{\perp},z)\,\,=\,\,  Q z ( 1 - z) \ko\,\,.
\eeq
$\s(r_{\perp},q^2_{\perp})$ is the cross section of the $\bar q q $ pair
with a transverse separation $r_{\perp}$ which scatter
off the target with momentum transfer
$q_{\perp}$. We first evaluate it  at
 $q_{\perp}$ = 0.

\subsection{$\s(r_{\perp},q^2_{\perp})$ at $q_{\perp}$ = 0}
The expression for
$\s(r_{\perp},q^2_{\perp})$ at $q_{\perp}$ = 0, was first written
down in Ref.[10]
(see Eq.(8) of this paper). It turns out that $\s$ can be
expressed through  the unintegrated parton density
 $\phi$ first introduced  in the BFKL papers[18]
and widely used in Ref.[2]. The relation of this
function to the Feynman diagrams and the gluon distribution can
be calculated using the following equation:
\beq
 \as (Q^2)\,\cdot\, x G(Q^2,x)\,\,=\,\,\int^{Q^2} \,d l^2_{\perp}\,\,
\as(l^2_{\perp})\,\,
\phi(l^2_{\perp}, x )\,\,.
\eeq
Using the above equation we reproduce the results of Ref.[10],
which reads
\beq
\s(r_{\perp},q^2_{\perp})\,\,=\,\,\frac{16 C_F}{N^2_c - 1}\,\,\pi^2
\int \phi(l^2_{\perp},x)\,\{\,\,1\,\,-\,\,e^{i\,\vec{l}_{\perp}
\vec{r}_{\perp}}\,\,\}\,\,\frac{\as(l^2_{\perp})}{2 \pi}\,
\frac{d^2 l_{\perp}}{l^2_{\perp}}\,\,,
\eeq
where $\phi \,\,=\,\,\frac{\pa x G (Q^2,x)}{\pa  Q^2}$ and
$C_F \,= \,\frac{{N_c}^2 - 1}{2N_c}$.
We evaluate this integral using Eqs.(3,4) and integrate over
the azimuthal angle.
  Introducing a new variable $\xi = r_{\perp} l_{\perp}$,
the integral can be written in  the form
\beq
\s(r_{\perp},q^2_{\perp})\,\,=\,\,\frac{16 C_F \as }{N^2_c - 1}\,\,\pi^2
\int_C \frac{d \o}{ 2 \pi i} \,\,g(\o)\,\,\g(\o)\,\,(r^2_{\perp})^{1 - \g(\o)}
\int^{\infty}_{0} d \,\,\xi\,\, \frac{1 - J_0(\xi)}{ (\xi)^{3 - 2 \g(\o)}}\,\,.
\eeq
 Evaluating the integral over $\zeta$ (see Ref.[19] {\bf 11.4.18})
we have
\beq
\s(r_{\perp},q^2_{\perp})\,\,=\,\,\frac{8 C_F \as }{N^2_c - 1}\,\,\pi^2
\int_C \frac{d \o}{ 2 \pi i} \,\,g(\o)\,\,\g(\o) \,\,
(\frac{r^2_{\perp}}{4})^{1 - \g(\o)}\,\,\,\frac{\G(\g(\o) \,\,
\G(1 - \g(\o))}{(\,\G(2 - \g(\o)\,)^2}\,\,.
\eeq
In the double log approximation of pQCD, where $\g(\o)\,\,\ll\,\,1$,
 the cross section  for $N_c \, = \, N_f$ = 3 reads
\beq
\s(r_{\perp},q^2_{\perp})\,\,=\,\,\frac{\as(\frac{4}{r^2_{\perp}})}{3}
\,\,\pi^2\,\,r^2_{\perp}\,\,
\(\,\, x G^{GLAP}( \frac{4}{r^2_{\perp}},x )\,\,\)\,\,.
\eeq
This result coincides with the value of the
cross section given in Refs.[6,20],
(if we neglect the factor 4 in the argument of the gluon
density). We checked that Eq.(2.16) of Ref.[7] also
leads to the same answer, unlike the value for $\s$ quoted
 in Ref.[7] (see Eq.(2.20)) which differs from ours by a factor of 2.

 We do not want to make use of the double log
approximation, which corresponds to the first term
in the expansion given in Eq.(5), and
prefer to use the complete series for $\g(\o)$. Indeed, all diagrams in
which an extra gluon goes from the bottom to the top as well as from
left to  right of the diagram of Fig.1 (see also Fig.2) do not produce
 an extra power of ln $\frac{1}{x}$. They only
give  corrections of the order $\as$ either to the anomalous
dimension, or to the coefficient
 function.  Calculating in the ln$\frac{1}{x}$ approximation,
considering $\frac{\as}{\o}\,\gg \as$ in the region of small $x$,
we can safely use the value of the anomalous dimension given in Eq.(5).
 From Eq.(14) one can directly obtain   the simple answer in another
 limiting case, namely,
when $\g(\o) \,\,\ra \,\frac{1}{2}$. This limit corresponds to the solution of
 the BFKL evolution equation[18] in the diffusion region when $\ln
 \frac{4}{r^2_{\perp}}\,\,\leq \,\sqrt{\as \ln \frac{1}{x}}$. Neglecting
 the deviation of $\g$ from $\frac{1}{2}$ we get
\beq
\s(r_{\perp},q^2_{\perp})\,\,=\,\,\frac{2 \as}{3}\,\,\pi^2\,\,r^2_{\perp}\,\,
\(\,\, x G^{BFKL}( \frac{4}{r^2_{\perp}},x )\,\,\)\,\,.
\eeq
In the intermediate range of $Q^2$ and $x$, we cannot obtain a
simple formula for $\s$ in terms of the
 gluon density. We stress that a difference of
a factor of two  in $\s$ does not mean that the cross section
 of vector meson production will change by a factor four.
The BFKL density behaves as $ r_{\perp} F(\ln r^2_{\perp},x)$
, and the integral over
 $r_{\perp}$ will be different for the double log approximation
 and for BFKL. We will consider these two cases and will
 integrate over $r_{\perp}$ to see how big  the difference is
 in the final answer.

\subsection{The amplitude of vector meson production at $q_{\perp}$ = 0}
We wish to calculate the amplitude for vector
meson production, using Eq.(9).
To do this we use the integral representation for the McDonald
function[19] and rewrite Eq.(10) in the form
\beq
\Psi^{\g^*} (r_{\perp},z)\,\,=\,\,\frac{Qz ( 1 - z)}{2}\,
\,\int^{\infty}_{0}
\,\frac{d t}{t}\,\,exp\(\, - \frac{1}{2}\,[ t\,+\,
\frac{ a^2\,r^2_{\perp}}{t}\,]\,\)\,\,.
\eeq
Substituting Eqs.(14,16) in Eq.(9) we obtain
\beq
M_f\,\,=\,\,\sqrt{N_f}\,\,\sum_{\lambda_1 \,\lambda_2}
 \int^1_0 \,d\,z\,\,\int\,\frac{d^2 r_{\perp}}{2 \pi}\,\,\,
\frac{Q z ( 1 - z)}{2}\,\int^{\infty}_{0}
\,\frac{d t}{t}\,\,exp\(\, - \frac{1}{2}\,[ t\,+\,
\frac{ a^2\,r^2_{\perp}}{t}\,]\,\)
\eeq
$$
\,\,\frac{8 C_F \as }{N^2_c - 1}\,\,\pi^2
\int_C \frac{d \o}{ 2 \pi i} \,\,g(\o)\,\,\g(\o) \,\,
(\frac{r^2_{\perp}}{4})^{1 - \g(\o)}\,\,\,\,\frac{\G(\g(\o) \,\,
\G(1 - \g(\o))}{(\,\G(2 - \g(\o)\,)^2}
\,\, {[\Psi^{V}(r_{\perp},z)]}^*\,\,.
$$
Anticipating that the typical value of $r_{\perp}$
in the integral will be small
  ($r_{\perp} \,\propto\,\frac{1}{Q}$),  we can replace the hadron
wave function by its value at $r_{\perp} $= 0. Introducing a new variable
 $\zeta\,\,=\,\,\frac{a^2 r^2_{\perp}}{2t}$, we integrate over
$r^2_{\perp}$ and t. It is easy to see that  the integral
corresponds to $\zeta
\,\simeq \,1$ and  $t \,\simeq \,2$. This means that the typical value
of $r^2_{\perp}$ is of the order of $\frac{4}{a^2} \,\ll\,1$.
This justifies our assumption that the wave function of a produced
hadron enters at small $r_{\perp}$.
\newpage
 The final general answer is
\beq
M_f\,\,=\,\,\sqrt{N_f}\,\,\sum_{\lambda_1 \,\lambda_2}
 \int^1_0 \,d\,z\,\,\,\,
\frac{Q z ( 1 - z)}{2}
\,\,\frac{8C_F \as}{N^2_c - 1}\,\,\pi^2
\int_C \frac{d \o}{ 2 \pi i} \,\,\,g(\o)
\eeq
$$
\,\, \frac{16}{a^4}\,\,
(a^2)^{\g(\o)}\,\,\G(1 + \g(\o)) \,\,
\G(1 - \g(\o))
\,\,{[\Psi^{V}(r_{\perp}= 0,z)]}^*\,\,.
$$
 In the case of the GLAP approach we can simplify Eq.(19) by
considering  $\g(\o) \,\,\ll\,\,1$, while for  BFKL  the
simplification occurs when $\g(\o) \,\ra\,
\frac{1}{2}$. One can see that there is a difference of $\pi/2$
in the amplitude,
but the final estimate is obtained only after integration over $z$.

We can approximate the integration over $z$
by putting $z = \frac{1}{2}$ in the ln$a^2$

dependence of the density.
This gives
\beq
M_f \,=\,C Q \,\,\as\(\, x G^{GLAP} ( a^2(z = \frac{1}{2}),x)\,\)
\,\,\int^1_0\, d\,z \,\,\frac{z(1-z)}{ a^4}  {[\Psi^V(0,z)]}^*\,\,,
\eeq
for the GLAP equations, while for the BFKL we get
\beq
M_f\,=\,C\,Q \,\as\,\,\frac{\pi}{2}\,
 \( x F^{BFKL}(\ln a^2(z=\frac{1}{2}),x)\,\)
\,\,\int^1_0\, d \,z \,\,\frac{z(1-z)}{ a^3}  {[\Psi^V(0,z)]}^*\,\,,
\eeq
where we have replaced the BFKL gluon distribution by $
x G^{BFKL}(a^2,x ) = \frac{r_{\perp}}{R}\cdot F(\ln r^2_{\perp},x)$
as  discussed above.
In the above equations we have replaced
all constants by the coefficient $C$.

The function $ \Psi^V(0,z)$ is just the function which has been
introduced in Ref.[7] (see Eq.(2.25) there).
Using the asymptotic
form of the z-dependence, namely $\Psi^V \,\propto z ( 1 - z)$ for
the case of massless quarks, we have:
\beq
M_f\,=\,C\,\frac{\as}{Q^3} \, \( x G^{GLAP}(\frac{Q^2}{4},x)\)\,\,,
\eeq
\beq
M_f\,=\,C\,\frac{\as}{Q^3} \,\frac{\pi^2}{8} \( x G^{BFKL}(\frac{Q^2}{4},x)\)
\,\,.
\eeq
Consequently, the difference in the value of the amplitude turns out
to be rather small, about 25\%. We note that the BFKL
equation gives the anomalous dimension close to $\frac{1}{2}$, which
only generates a
 $\frac{1}{Q^2}$ behaviour of the amplitude, while the GLAP equations give
$\frac{1}{Q^3}$ for the same value. In the nonrelativistic
case the difference
is $\pi/2$ in favour of the BFKL equation.

\subsection{The $b_{\perp}$ dependence of the amplitude}
To deal with the SC we need to know  the amplitude not only at
$q_{\perp}$=0, but also at all values of momentum transfer,
so that we can calculate
 the profile function of the amplitude in impact
parameter space. The gluon density has a week dependence on $q_{\perp}$
 at small values of $q_{\perp}$,
 both in the double log approximation (see Ref.[2])
as well as for the BFKL approach (see Ref.[20]).
Therefore, the $q_{\perp}$-dependences come from
the form factor of the $\bar q q$ pair with a transverse separation
$r_{\perp}$, and the form factor of the proton target.

 The target form factor is not
 treated theoretically in pQCD, and  that for our
purpose we assume
the exponential parameterization for it will suffice, namely
$$F_p ( q^2_{\perp})\,\,=\,\,e^{- \frac{B}{4} \,\,q^2_{\perp}}\,\,.$$
If we put the Pomeron slope $\a'$=0
the slope $B$ can be extracted from the experimental data on hadron-hadron
collisions.  Namely,
 $B$ = $B^{pp}_{el}(\a' = 0 )$, where $B_{el}$ is the slope of the differential
cross section of proton-proton collision. For the numerical estimates we
use  phenomenological information  to extract the value of $B$ [21].
The value of $B$ defined in this way, is very close to
the one obtained from the proton electromagnetic form factor.

The form factor of the $\bar q q $ pair with a transverse separation
$r_{\perp}$ is equal to
\beq
F_{\bar q q} ( q^2_{\perp})\,\,=\,\,\Psi^{i}_{\bar q q} \(
\frac{({\vec{k}}_{1\perp}\,\,-\,\,{\vec{k}}_{2\perp})\,\cdot\,
{\vec{r}}_{\perp}}{2} \) \,\Psi^{f*}_{\bar q q} \(
\frac{({\vec{k'}}_{1\perp}\,\,-\,\,{\vec{k'}}_{2\perp})\,\cdot\,
{\vec{r}}_{\perp}}{2} \)\,\,,
\eeq
where $k_i$ ($k'_i$) denotes the momentum of the $i$ quark before and
 after the collision. Each of the wave functions has an
exponential form, and
a simple sum of different attachments of gluon lines to quark lines
 gives
\beq
F_{\bar q q} ( q^2_{\perp})\,\,=\,\,e^{ i \frac{ {\vec{q}}_{\perp}\,\cdot \,
{\vec{r}}_{\perp}}{2}}\,\,\{\,\,1\,\,-\,\,e^{i\,\vec{l}_{\perp}\,\cdot \,
\vec{r}_{\perp}}\,\,\}\,\,.
\eeq
We have absorbed the last factor in the expression for the cross section,
while the first one gives the $q_{\perp}$ dependence of the $\bar q q $
 form factor, which after integration over the azimuthal angle
has the form
\beq
F_{\bar q q} (
q^2_{\perp})\,\,=\,\,J_0(\frac{q_{\perp}\,\,\,r_{\perp}}{2})\,\,.
\eeq
To calculate further, we need the profile function in $b_{\perp}$ space,
which is defined as
\beq
S(b^2_{\perp})\,\,=\,\,\frac{1}{4\pi^2} \,\,\int \,d^2  q_{\perp}\,\,
e^{i {\vec{b}}_{\perp}\,\cdot\,{\vec{q}}_{\perp}}\,\,\,F_p ( q^2_{\perp})\,\,
 F_{\bar q q} ( q^2_{\perp})\,\,.
\eeq
An exact computation gives
\beq
S(b^2_{\perp})\,\,=\,\,\frac{1}{\pi B}\,\,I_0
\(\frac{b_{\perp}\,\,r_{\perp}}{B}
\)\,\,e^{-\,\,\frac{b^2_{\perp}\,\,+\,\,\frac{r^2_{\perp}}{4}}{B}}\,\,.
\eeq
To simplify the calculation we replace the above function by
\beq
S(b^2_{\perp})\,\,=\,\,\frac{1}{\pi B'}\,\,e^{- \frac{b^2_{\perp}}{B'}}\,\,,
\eeq
where
$$ B'\,\,=\,\,B\,\(\,1\,\,+\,\,\frac{r^2_{\perp}}{4 B}\,\)\,\,\simeq \,\,
B\,\(\,1\,\,+\,\,\frac{1}{a^2 B}\,\)\,\,.$$
Eq.(29), in the small
 $b^2_{\perp}$ expansion, has the same radius as in Eq.(28).

\section{Shadowing corrections to diffractive
\newline
leptoproduction of vector mesons }
\subsection{Percolation of a  $\bar q q $-pair through the target.}
To calculate the shadowing corrections we follow the procedure
suggested in Refs.[9,10,13,21].
Namely, we replace $\s(r_{\perp},q^2_{\perp} = 0 )$ in Eq.(9) by
\beq
\s^{SC}(r_{\perp})\,\,=\,\,2\,\,\int d^2 b_{\perp} \,\( \,\,1\,\,-\,\,
e^{-\,\frac{1}{2}\,\,\s(r_{\perp},q^2_{\perp} = 0 )\,\,S(b^2_{\perp})}\,\,\)
\,\,.
\eeq
The above formula is a solution of the s-channel unitarity relation
\beq
2\,\,Im\,a(s,b_{\perp})\,\,=\,\,|a(s,b_{\perp})|^2 \,\,+\,\,G_{in}(s,b_{\perp})
\,\,,
\eeq
where $a$ denotes the elastic amplitude for the $\bar q q $ pair with a
transverse separation
 $r_{\perp}$, and $G_{in}$  is the contribution of all the
inelastic processes. The inelastic cross section is equal to
\beq
\s_{in}\,\,=\,\,\int \,d^2 \,b_{\perp}\,\,G_{in} ( s, b_{\perp})
\,\,=\,\,\int d^2 b_{\perp} \,\,\( \,\,1\,\,-\,\,
e^{-\,\,\s(r_{\perp},q^2_{\perp} = 0 )\,\,S(b^2_{\perp})}\,\,\)\,\,.
\eeq
We assume that
the form of the final state is a uniform parton distribution that
follows from the QCD evolution equations. Note that we neglect
the contribution of all
diffraction dissociation processes to the inelastic final state
(in particular to the ``fan" diagrams which give an  important
contribution[2]),
as well as  diffraction dissociation in the region of small
masses[21], which cannot
 be presented as a decomposition of the $\bar q q$ wave function.
We  evaluate this input  hypothesis in the next section.

In the language of Feynman diagrams, Eq.(32)
sums all diagrams of Fig.3 in which
 the $\bar q q $ pair rescatters with the target, and exchanges
 ``ladder" diagrams, each of which corresponds to the
gluon density.
This sum has already been performed by Mueller[9],
and we will only comment on
 how one can obtain the result, without going into details.

The  simplest way is to consider the inelastic cross section
(see Refs.[2,9]), which has a straightforward
interpretation through the parton wave function of the hadron (see
Fig.4), as all partons are produced on the mass shell.
In leading ln$\frac{1}{x}$ approximation we have two
orderings in time (see Fig.4):
\newline
\newline
1. The time of emission of each ``ladder" by the fast $\bar q q $ pair,
which should obey the obvious ordering for $n $ produced ``ladders"
\beq
t_1\,\,>\,\,t_2\,\,...\,\,>\,\,t_{i}\,\,>\,\,t_{i + 1}\,\,...>\,\,
t_{n}\,\,.
\eeq
2. Each additional ``ladder"  which should live for a shorter time than the
previous one. This gives a second ordering
\beq
t_1 - t'_1\,\,>\,\,t_2 - t'_2\,\,>\,\,...\,\,> \,\,t_i - t'_i\,\,>\,\,...
\,\,>\,\,t_n - t'_n\,\,.
\eeq
Each ``ladder" in the leading log approximation has the same
functional dependence on $t_i - t'_i$, which we denote
as $\s(t - t',r^2_{\perp})$. This
fact allows us to carry out the integration over $t_i$ and $t_i - t'_i$,
which gives
for $n$ emitted cascades
\beq
\s_n \,\,=\,\,\frac{1}{(n!)^2} \int^t d t_1 \int^{t - t'} d ( t_1 - t'_1) \,
\,\s^n(t_1 - t'_1,r^2_{\perp})\,\,.
\eeq
For the case of the $\bar q q $ pair,  both last integrations are not
logarithmic, and in the LLA
we can safely replace the above
integral with
\beq
\s_n\,\,=\,\,\,\frac{1}{(n!)^2}  \,
\,\s^n(t - t',r^2_{\perp})\,\,,
\eeq
$t - t'$ is of the order of $\frac{1}{q_{\parallel}}$ due to the uncertainty
 principle, where
 $q_{\parallel}\,\,=\,\,\frac{Q^2\,\,+\,\,m^2_V}{s}$.
One $n!$ is compensated by the number of possible
diagrams, since the order of the "ladders" are not fixed.
 Therefore, the contribution of
the $n$th ``ladder" exchange gives
\beq
\s_n\,\,=\,\,\,\frac{1}{n!}  \,
\,\s^n( x,r^2_{\perp})\,\,.
\eeq
Applying the AGK cutting rules[22], we reconstruct the total cross
section which results in Eq.(32).

\subsection{Damping factor for diffractive production of vector meson}
Substituting $\s^{SC}$ of Eq.(32) in Eq.(9) and using
$\s$ in the form of Eq.(14)
and the representation (17) for the wave function of the photon,
we can easily
estimate the general term of the expansion with respect to the power of $\s$.
It has the form
\beq
M^{n}_f = C \int^1_0 \,d\,z \frac{Q z ( 1 - z )}{2}\,\,
\int^{\infty}_{0}\,\,\frac{d t}{t} \,exp \(- \frac{1}{2}\,[t\,\,+\,\,
\frac{a^2\,\,r^2_{\perp}}{t}\,\,]\,\)\,\,\frac{( - 1 )^{n- 1}}{n!}
\,\(\frac{\as\,4 C_F \pi^2}{N^2_c - 1}\)^n \,\,
\eeq
$$
\prod^n_{i}\,\,\int_{C_i} \,\,\frac{d\,\o_i}{2 \pi i}\,\,g_i (\o_i)\,
\,\,\frac{\G(1 + \g(\o_i))\,\,\G(1 - \g(\o_i))}{( \G(2 - \g(\o_i))^2}
$$
$$
\,\,\(\frac{r^2_{\perp}}{4}\)^{n \,\,-\,\,\sum^n_i \,\g(\o_i)}
\,\,\,{[\Psi^V(r_{\perp} = 0, z)]}^*\,\,\,\int \,d^2 b_{\perp}\,\,
S^n(b^2_{\perp})\,\,.
$$
Taking the integrals over $r_{\perp}$, $t$ and $b_{\perp}$ we obtain
\beq
M^{n}_f = C \,\,B'\,\, \int^1_0 \,d\,z \frac{Q z ( 1 - z )}{2\,a^2}\,\,
\frac{( - 1 )^{n - 1}}{n\,\,n!}
\(\frac{\as\, C_F \pi}{B'(N^2_c - 1)}\)^n
\eeq
$$
\prod^n_{i}\,\,\int_{C_i} \,\,\frac{d\,\o_i}{2 \pi i}\,\,g_i (\o_i)\,\,
\,\frac{\G(1 + \g(\o_i))\,\,\G(1 - \g(\o_i))}{( \G(2 - \g(\o_i))^2}
$$
$$
\,\,
\,\,\G^2( 1 + n - \sum^n_i \g(\o_i))\,\,\(\,\frac{4}{a^2}\,\)^n\,\,(a^2)^{\sum
\g(\o_i)}
\,\,{[\Psi^V(r_{\perp} = 0, z)]}^*\,\,.
$$
In the double log approximation of pQCD, $\g(\o_i)\,\,\ll\,\,1$  and we
 derive a very simple formula (neglecting $\g$). Taking
the integral over $\o_i$
we have
\beq
M^{n}_f = C \,\,B'\,\, \int^1_0 \,d\,z \frac{Q z ( 1 - z )}{2\,a^2}\,\,
\frac{( - 1 )^{n - 1}n!}{n}
\eeq
$$
\,\(\frac{ C_F \pi}{B'(N^2_c - 1)}\,\,\frac{4}{a^2} \,\,\as\,\,
x G(a^2,x) \)^n
\,\,{[\Psi^V(r_{\perp} = 0, z)]}^*\,\,.
$$
Finally for $M_f$ we have
\beq
M_f\,\,=\,\,C\,\, B'\,\,\sum^{\infty}_{n = 1} \int^1_0 \,d\,z
 \frac{Q z ( 1 - z )}{2\,a^2}\,\,(-1)^{n-1}(n - 1 )!\,\,
\eeq
$$
\(\frac{ C_F \pi}{B'(N^2_c - 1)}\,\,\frac{4}{a^2} \,\,\as\,\,x G(a^2,x) \)^n
\,\,{[\Psi^V(r_{\perp} = 0, z)]}^*\,\,.
$$
The above series can be written in terms of
the analytical function $E_1$ (see Ref.[19]), namely
\beq
M_f\,\,=\,\,C\,\, B'\,\,\int^1_0 \,d\,z
 \frac{Q z ( 1 - z )}{2\,a^2}\,\,E_1 (\,\frac{1}{\kappa}\,)\,
e^{\frac{1}{\kappa}}
\,\,{[\Psi^V(r_{\perp} = 0, z)]}^*\,\,,
\eeq
where (for $N_c$= 3)
\beq
\kappa_q\,\,=\,\,\frac{2}{3}\,\cdot\,\frac{\as \pi}{B' \,\,a^2} \,\,
x G( a^2, x)\,\,.
\eeq
Using the above equation we have for the damping factor
\beq
D^2\,\,=\,\,\frac{ \{\int^1_0 \,d\,z
 \frac{Q z ( 1 - z )}{2\,a^2}\,\,E_1 (\,\frac{1}{\kappa_q}\,)\,
e^{\frac{1}{\kappa_q}}
\,\,\Psi^V(r_{\perp} = 0, z)\}^2}{\{\int^1_0 \,d\,z
 \frac{Q z ( 1 - z )}{2\,a^2}\,\,\kappa_q
\,\,\Psi^V(r_{\perp}= 0,z )\}^2}\,\,.
\eeq
For the case of  heavy mesons we can use the nonrelativistic approach
$ z \ra \frac{1}{2}$,   and the above formula has the very simple form
\beq
D^2\,\,=\,\,\frac{\{ E_1(\frac{1}{\kappa_q})\,e^{\frac{1}{\kappa_q}}\}^2}
{\kappa^2_q}\,\,.
\eeq
The behaviour of the damping factor can be easily found using the
 well known property of $E_1$. Namely,
at $\kappa_q \,\ll\,1$, we have $D^2 \ra 1$ while at $\kappa_q\,\gg\,1$
the damping factor vanishes as
 $D^2 \propto \frac{\ln^2 \kappa_q}{\kappa^2_q}$. The behaviour of
the damping factor as a function of $\kappa$ is given in Fig.5.
\section{The relation between
DIS diffractive production of vector mesons
and $F_2(Q^2,x)$.}
The expression for the $F_2$ structure function was derived by
Mueller[9] and in our notation has the following form
\beq
\Sigma(Q^2,x)\,\,=\,\,\int^1_0 \,d\,z \,\int\,\frac{d^2 r_{\perp}}{2 \pi}
\,\,\Psi^{\g^*}_{\perp}(Q^2,r_{\perp},z)\,\,\s^{SC}( r_{\perp})\,
\,{[\Psi^{\g^*}_{\perp}(Q^2,r_{\perp},z)]}^*\,\,,
\eeq
where

\beq
\Sigma(Q^2,x)\,\,=\,\,\frac{F_2(Q^2,x)}{\VEV{e^2}}\,\,.
\eeq
$\VEV{e^2}$ denotes the average quark charge, which is equal to
$\frac{2}{9}$ ($\frac{5}{18}$) for three
(four) active flavors respectively.

$\Psi^{\g^*}_{\perp}$ represents the wave function of the virtual
photon with transverse polarization.
$\Psi^{\g^*}_{\perp}$, as needed for our calculations,
has been given in Ref.[9] (see also Refs.[17,23])
i.e.
\beq
\Psi^{\g^*}_{\perp}(Q^2,r_{\perp},z)\,\,=\,\,a\,K_1( a r_{\perp})\,\,.
\eeq
However, as was shown in Ref.[9], within the LLA
of pQCD, after
 integration over $z$ in Eq.(46),
 we can safely replace this function by $\frac{1}{r^2_{\perp}}$.
 Finally  Mueller's result reads
 (for $N_c = N_f $= 3)
\footnote{We thank A. Mueller for a discussion with us on this problem.}
\beq
\Sigma(Q^2,x)\,\,=\,\,\frac{4}{ \pi^2}\,\,\int\,
\frac{d^2 \,b_{\perp}}{ \pi}\int^{\infty}_{\frac{4}{Q^2}}
\,\frac{d^2\,r_{\perp}}{ \pi}\,\,
\frac{1}{r^4_{\perp}}
\,\,\s^{SC}( r_{\perp})\,\,.
\eeq
Comparing this equation with the general Eq.(9) we conclude that
\beq
M_f\,\,=\,\,\sqrt{N_f}\,\,\frac{4}{ \pi^2}\,\,\int^1_0\,d\,z
\,\,\int^\infty_\frac{4}{Q^2}\,\,\frac{d^2 \,r_{\perp}}{\pi} \,\,
{[\Psi^V (r_{\perp},z)]}^*
\,\,\frac{\pa \Sigma(\frac{4}{r^2_{\perp}},x)}{\pa \ln \frac{1}{r^2_{\perp}}}
\,\,.
\eeq
Using the notations of Ref.[7] and our normalization of the amplitude,
we can rewrite the expression for the amplitude in the form
\beq
f ( \g^* \,\ra\, V )\,\,=\,\,\frac{4 e f_V}{\sqrt{2}}\,\int^1_0\,d\,z
\,Q\,z\,( 1 - z)\,\int^\infty_\frac{4}{Q^2}\,r^3_{\perp}\,d\,r_{\perp}\,\,K_0 (
a\,r_{\perp})
\eeq
$$
\,\,\frac{\pa F^{exp}_2
(\frac{1}{r^2_{\perp}},x)}{\pa\,\ln \frac{1}{\,r^2_{\perp}}}
\,\,\frac{\frac{1}{2}\,\,\varphi^V(z)}{\int \,d z\, \varphi^V (z)}\,\,,
$$
where $\varphi^V (z)\,=\,\Psi^V(r_{\perp} = 0, z)$.

The atractive feature of the above formula is that it takes into
account all shadowing corrections
 in the LLA of pQCD, and  expresses
 the cross section for vector meson production directly through the
experimental observable. We note that without
SC this formula is trivial and follows directly from the GLAP evolution
 equations in the LLA of pQCD (see, for example, Ref.[24]).
The shortcoming of  this formula
 is obvious: we cannot perform the integration over
 $r_{\perp}$ as has been done in Eq.(44). We think that this formula
will be especially
  useful for reactions with a nuclear target, since it absorbs all
SC which are essential in this case.

Using Eq.(51) we can rewrite the expression for the damping factor in the form
\beq
D^2\,\,=\,\,\frac{\{\int^1_0\,d\,z
\,Q\,z\,( 1 - z)\,\int\,r^3_{\perp}\,d\,r_{\perp}\,\,K_0 ( a\,r_{\perp})
\,\,\frac{\pa F^{exp}_2
(\frac{1}{r^2_{\perp}},x)}{\pa\,\ln \,\frac{1}{r^2_{\perp}}}
\,\,\frac{\frac{1}{2}\,\,\varphi^V(z)}{\int^1_0 \,d z\, \varphi^V (z)}\}^2}
{\{\int^1_0\,d\,z
\,Q\,z\,( 1 - z)\,\int\,r^3_{\perp}\,d\,r_{\perp}\,\,K_0 ( a\,r_{\perp})
\,\,\frac{\pa F^{GLAP}_2
(\frac{1}{r^2_{\perp}},x)}{\pa\,\ln \frac{1}{\,r^2_{\perp}}}
\,\,\frac{\frac{1}{2}\,\,\varphi^V(z)}{\int \,d z\, \varphi^V (z)}\}^2}\,\,.
\eeq
In the case of a nuclear target, we have
$$\frac{\pa F^{GLAP}_2(A)}{\pa ln\frac{1}{ r^2_{\perp}}} \,\,=
\,\,A \frac{\pa F^{GLAP}_2(N)}{\pa ln \frac{1}{r^2_{\perp}}}\,\,,$$
where $F_2(N)$ is the structure function for a DIS with
a nucleon. The last remark makes the whole expression  of Eq.(52) useful
for a nuclear target, since all values that enter the formula can be
 measured experimentally. In the case of a nucleon target,
we have to rely on
the solution of the GLAP evolution equations to estimate the denominator.

\section{Shadowing corrections for the gluon distribution}

 To calculate the damping factor using Eq.(44) we need to calculate
the SC for the gluon distribution. We anticipate that these
shadowing correction will be large as
\newline
\newline
1. The
cross section of the interaction of two gluons with a transverse separation
$r_{\perp}$ is bigger than for $\bar q q $ pair and is equal to[9]
\beq
\s^{GG}\,\,=\,\,\frac{3 \as}{4}\,\,\pi^2\,\,
r^2_{\perp}\,\,x G(\frac{4}{r^2_{\perp}},x)\,\,.
\eeq
2.  Each gluon in the parton cascade can interact with the target.
Such interactions generate the so called ``fan" diagrams (see Fig.6),
which are essential in the region
 of small $x$, and correspond to the interaction of
a fast $GG$ pair[2] (see Fig.3).

In this paper we chose the following strategy of how to calculate
 the SC to the gluon distribution. First, we consider the SC for
the fast gluon  and discuss the contribution of the large distances.
After that, we will discuss
the contributions of the ``fan" diagrams.

 The SC for the fast $GG$ pair has been
calculated by Mueller[9],
and in our notation the gluon distribution has the form
( $N_c = N_f$ = 3)
\beq
x G(Q^2,x)\,\,=\,\,\frac{4}{ \pi^2}\,\,\int^1_x \,\frac{d x'}{x'}
\int\,\,
\frac{d^2 \,b_{\perp}}{\pi}\int^{\infty}_{\frac{4}{Q^2}}
\,\frac{d^2\,r_{\perp}}{\pi}\,\,
\frac{1}{r^4_{\perp}}
\,\,2\,\{1\,\,-\,\,e^{ -\,\frac{1}{2}\,\s^{GG}(r^2_{\perp},x')\,
S(b^2_{\perp})}\}\,\,.
\eeq
To understand the physical meaning of this equation it is  instructive
to write down the
\newpage
evolution equation for the gluon density.
Indeed,
\beq
\frac{\pa^2 x G(Q^2,x)}{\pa \ln(1/x) \,\pa \ln Q^2}\,\,=
\eeq
$$\,
\frac{N_C \as}{\pi}\,\,x \,G^{GLAP}(Q^2,x)\,\,+\,\,\frac{1}{\pi^2}\,
\sum_{k=1} \frac{(- 1)^k}{k k!}\,\,
\frac{1}{( B' Q^2)^{ (k - 1)}}\,\,
\( \frac{\pi C_A \as x G^{GLAP} (Q^2,x)}{2}\)^{k + 1}\,\,.
$$
The first term corresponds to the usual GLAP equations, while the second
one takes into account the SC.
 It should be stressed that the term with $k$ = 1 (if
treated as an equation as in Ref.[9]),
is the same term that appears
 in the nonlinear GLR equation which sums the ``fan" diagrams[2].
 This term has been calculated using quite
a different technique[25,26].
The coefficient in front of the other terms, reflects the fact that all
 correlations between the gluons have been neglected, despite the fact
that the gluons are uniformly distributed in the
target disc with a radius $R^2 = B'$.

 Mueller's formula is not a nonlinear equation, it is the analogue
of the Glauber formula for the interaction with a nucleus,
which gives us the possibility
 to calculate the shadowing corrections using the solution of the
 GLAP evolution equations. Hence, this formula should be used as an
input to obtain the complete
  effect of the SC,
 for the more complicated evolution equations, such as
GLR[2], or the generalized evolution equations[27].

Calculating the sequent iterations of Mueller's formula
gives us a practical way to estimate the value of the SC from more
complicated Feynman diagrams, that have to be taken into account (diagrams
such as in Fig.3). Certainly, the iteration of Muller's formula is not the
most efficient way
 to calculate the SC correction in the region  of extremely
  small $x$, but,  it
 gives a systematic way to estimate the value of the SC due to gluon screening
 and it could be a reliable procedure in the HERA kinematic region where
 $x$ is not too small.

Making use of the explicit form of $S(b_{\perp})$ (see Eq.(29)), we take the
integral over $b_{\perp}$ and obtain
\beq
x G(Q^2,x)\,\,=\,\,\frac{2}{ \pi^2}\,\,\int^1_x \,\frac{d x'}{x'}
\,\,\int^{\infty}_{\frac{1}{Q^2}}
\,\frac{d^2\,r'_{\perp}}{\pi}\,\,
\frac{B'}{r'^4_{\perp}}
\,\,\{ \,C\,\,+\,\,\ln \kappa_G(x',r'^2_{\perp})\,\,+\,\
,E_1 ( \kappa_G(x',r'^2_{\perp})\,\,
\}\,\,,
\eeq
where $C$ is the Euler constant and $E_1$ is the exponential integral
 (see Ref.[19] {\bf 5.1.11}) and
\beq
\kappa_G(x',r'^2_{\perp})\,\,=\,\,\frac{3 \as \pi r'^2_{\perp}}{2  B'}\,\,
x' G^{GLAP}(\frac{1}{r'^2_{\perp}}, x'))\,\,.
\eeq
In our numerical estimates which will be presented in the next section
 we use this formula
to calculate the SC due to gluon screening. We also make the second iteration
 of Eq.(56) substituting the result of the first iteration back
into Eq.(56) in place of
$ x G^{GLAP}$. In this way we take into account the first contribution of the
``fan" diagrams (see Fig.7). We will discuss the result of these calculation
 in the next section.

\section {Numerics and comparison with experiment}
\subsection {Vector meson production in DIS without SC}
  We show in Fig.8a the dependence
of $\sigma(\g^*p \ra  Vp)$
on $x$ with no SC for some typical values of $Q^2$.

 As we have seen in section 2, the pQCD matrix element for DIS production
of vector mesons is given as a convolution of
$\Psi^{\g^*} (Q^2,r_{\perp},z)$,
$\Psi^V (r_{\perp},z)$ and $\sigma(r_{\perp},{q_{\perp}}^2)$
where we are able to calculate the above when both $\g^*$ and V are
longitudinally polarized. The specifications of
$\Psi^{\g^*}$, $\Psi^V$ and $\sigma$ used in our calculations are as follow:
\newline
\newline
1. We use the commonly utilized form for $\Psi^{\g^*}$, e.g. our Eq.(10).
\newline
\newline
2. For $\Psi^V$ we use the two forms which were also used in Ref.[7].
The asymptotic form[28] denoted AS and the form suggested by Chernyak
and Zhitnitsky[29] denoted CZ. We note that Braun and Halperin[30]
have cast some doubts on the CZ form and have suggested a modified wave
function not that different from the AS form.
The difference between our results with the AS and the CZ inputs,
as seen in Fig.8a, defines a theoretical margin of error due to the
ambiguity in our knowledge of $\Psi^V$.
This should be compared later on with the margin of difference introduced
due to the SC.
\newline
\newline
3. The ${\bar q}q$ cross section is given in the LLA utilizing $xG^{GLAP}$,
e.g. our Eq.(15), with the GRV parameterization[31].
In our opinion GRV is suitable for our calculations as they chose a
small scale and as such their result for small $x$ are close to the LLA,
enabling  a reasonably consistent calculation.

 A semiquantitative calculation of
$\sigma(\g^*p  \ra  \rho p)$ has been presented by Brodsky et.al.[7]
and is compared to the low energy NMC data[32] at $x$ = 0.06 and
$Q^2$ = 10 $GeV^2$.
We note a difference between the approximation used in Ref.[7] and ours.
When calculating the cross section shown in Fig.8a we take into account
a $z$ dependence of $\sigma(r_{\perp},{q_{\perp}}^2)$ introduced after the
$r_{\perp}$ integration. Brodsky et.al. neglect this $z$ dependence and as
a result their calculation with the AS or CZ wave functions have
a fixed ratio of $(\frac{3}{5})^2$. The equivalent ratio in our calculation
depends on $x$ and $Q^2$ as can be seen in Fig.8a. Note that in our
integration we have a cutoff $a^2 = Q^2 z > {Q_0}^2 = 0.4 {GeV}^2$.
The relatively strong dependence of the above ratio on $x$ and $Q^2$
shows that the LLA used to obtain Eq.(19) does not work too well in the
kinematical domain under investigation.
We shall return to this point in section 7.
Regardless of the above, the calculation of Ref.[7] and ours at
$x$ = 0.06 and $Q^2 = 10 GeV^2$ are mutually compatible and are
comparable to the NMC data[32].

\subsection{Vector meson production in DIS with SC}

 Fig 8a should be compared with Fig 8b where we present our results for
$\sigma(\g^* p \ra  \rho p)$
with SC as defined in section 3. Inspection of Fig.8 shows that the
reduction of the cross sections due to SC is bigger than the margin of
theoretical error shown by the difference between the AS and CZ calculations
of Fig.8a. As such SC are demonstrated to be an important ingredient of
the pQCD calculation and should not be ignored.
The comparison between the screened and non screened calculations of
$\sigma(\g^* p \ra  \rho p)$
are further shown in Fig.9 as a function of $Q^2$ for c.m. energies of
$\sqrt{s}$ = 50, 100 and 150 GeV. As expected, the difference between the
screened and non screened results gets smaller when $Q^2$ is increased
and becomes bigger when $\sqrt{s}$ is increased.

 Recently, some ZEUS data on $\sigma(\g^* p \ra  \rho p)$ became
available[33], and when combined with the low energy NMC data[32]
provide a first qualitative examination of the pQCD results.
\newline
\newline
1. The ZEUS differential cross section slope is found to be
$B = 5.1 +1.2 -0.9 \pm  1.0 GeV^{-2}$.
This B value is about a half of the slope found in elastic $\rho^0$
photoproduction and is in excellent agreement with our input[13,21].
\newline
\newline
2. ZEUS reports a $Q^{-n}$ dependence of $\s(\g^* p \ra  \rho p)$
where $n = 4.2 \pm 0.8 +1.4 -0.5$. This should be compared with the
(asymptotic) prediction of Brodsky et.al.[7] of n = 6.
In our calculations n is a function of $x$ and $Q^2$
approaching n = 6 when the SC are diminished.
Note that $ n \ra 2$ when $ln \frac{1}{x}$ is big enough,
actually well above the presently available HERA kinematics.
In the low $Q^2$ domain investigated by ZEUS,
$7 \leq Q^2 \leq 25 GeV^2$, our calculations result in n = 5.3 for AS
and n = 4.8 for CZ. Clearly the present experimental errors are too big
for a definite conclusion. Note, also, that the ZEUS results are obtained
for $\s = \s_L + \s_T$ with $R = \frac{\s_L}{\s_T} = 1.5 +2.8 -0.6$,
whereas the pQCD analysis applies to $\s_L$ only.
\newline
\newline
3. The dependence of the ZEUS-NMC cross sections on $x$ is compared with the
pQCD results in Fig 10 for $x \leq 0.01$. The unscreened results calculated
by ZEUS according to Ref.[7] are given by the shaded area,
whereas the screened
results are given by the two solid lines denoted CZ and AS.
At this time the
experimental errors are sufficiently large so as to accommodate both the
screened
and unscreened options. However, it is encouraging to note that a luminosity
increase by a factor of 2-3 will enable a data based discrimination between
the theoretical models. Note that the theoretical predictions on Fig.10 are
obtained after a multiplication by R = 1.5, to account for the observed
$\s_T$ as well.

 Fig.11 shows the contours of the damping factor $D^2$ (defined in Eq.(1)
and evaluated in Eqs.(44,45)) as a function of
$\sqrt{s}$ and $Q^2$ for $\g^* p \ra J/\psi  p$.
Note that for heavy quarks the dependence on $\Psi^V$ is eliminated in
the nonrelativistic limit.
Indeed, the difference between the AS and CZ results was small enough to
be neglected.
Experimental data
on this reaction in the HERA energy range is now available[34]
with $Q^2 \leq 4 GeV^2$. This data together with lower energy data is
shown in Fig.12 together with some theoretical predictions.
The shaded area corresponds to Ryskin's pQCD calculations[5]
and our corrections to Ryskin's predictions are
presented by the two solid lines.
Note that ours is an approximate
assessment due to the following uncertainties:
\newline
\newline
1. Ryskin  uses the BFKL kernel whereas we  use a GLAP kernel.
\newline
\newline
2. The shaded area in Fig.11 corresponds to the ZEUS experimental
values for $xG(x,Q^2)$ inserted into Ryskin's calculations.
Our calculation of $D^2$ is based on the GRV parameterization[31].
\newline
\newline
3. Ryskin uses a non relativistic quark model wave function for the
vector meson, whereas we assume the relativistic wave
functions which were also used  in Ref[7]. This difference results in
a normalization ambiguity between the two approximations which has not as yet
been resolved.
\newline
\newline
Regardless of the above, an
inspection of Figs.11 and 12
shows that if a pQCD
calculation, with no SC, yields an energy dependence of $s^{2 \Delta}$ for
$\s(\g^* p \ra J/\psi p)$, the imposition of SC will reduce $\Delta$ by
approximately 0.08 for the $Q^2$ range of Ref.[34]. For real photoproduction,
$Q^2$ = 0, the reduction in $\Delta$ will be 0.11.
\subsection{The damping factor with SC for the gluon distribution}
Fig.13 shows the damping factor for
$J/\psi$  production
 calculated using Eq.(56) for the gluon
distribution and Eq.(9) for the vector production amplitude.
As has been discussed,
we anticipate a bigger influence of
the small distances in the gluon damping than we have observed
in the the quark sector with SC.
The damping factor in the
gluon sector is calculated to be
\beq
D_G^2\,\,=\,\,\{ \,\frac{ xG(a^2(z = \frac{1}{2}),x)^{[Eq.(56)]} +
xG^{GRV}(Q^2_0,
x)}{\frac{N_c \as}{\pi} \,\int^1_x \,\frac{d x'}{x'} \int^{a^2}_{Q^2_0}\,
\frac{d Q'^2}{ Q'^2} \,x' G^{GRV}( Q'^2, x') \,\,+\,\, x G^{GRV} (Q^2_0,x)}\,
\}^2\,\,,
\eeq
where $Q^2_0$ = 0.4 $GeV^2$.
This should be compared with Eq.(44) in the quark sector.
In Eq.(56) we have integrated over $r^2_{\perp}$ from
$\frac{1}{a^2}$ to $\frac{1}{Q^2_0}$.
Note that the denominator of Eq.(56)is the Born Approximation
of the numerator.Since we have seen that the LLA
for the GLAP evolution equation does not work well in the kinematic region of
interest, we also use this approximation  for the denominator
in our calculation as the SC
is known only in LLA. We also demand that the initial condition for the gluon
distribution is the same for SC and without them.
Our reasoning is that we look on
the initial condition as if it has been taken from the experiment.
Therefore, the result of the calculation can be read in the following way:
 the suppression of $J/\psi$ production
is given by $D_G^2$ due to the
screening  at distances of
$\frac{1}{Q^2_0}\,\,>\,r^2_{\perp}\,>\,\frac{4}{Q^2}$.
One can extract from Figs.12 and 13
that the SC corrections in the gluon sector, are
as large as in the quark sector.
This fact could be anticipated because the
$\kappa_G$ is $\frac{9}{4}$ times bigger than $\kappa_q$. Additional
 integration over $x$ in Eq.(56) suppresses the gluon screening, but
  not to an extent that
will make it smaller than SC in the quark sector.

The ratio $R\,=\,\frac{D_G^2 (\,\,second\,\,\,\, iteration\,\,)}
{D_G^2 (\,\,first\,\,\,\, iteration\,\,)}$ is shown in
Fig.14. $R$ is the quantitative
 measure of the correction to the damping factor due to the second iteration
of Eq.(56) which takes into account the first correction from the ``fan"
diagrams (see Fig.7). One can see that this correction is rather big
 and we plan to study them in a more comprehensive way in
a future publication.

\section{Conclusions}
1. The formulae for shadowing corrections for vector meson production in
DIS have been obtained whihin the framework of the GLAP
evolution equations in the low $x$ region. It is shown that
 the rescattering of the quark is concentrated at small distances ( $r_{\perp}
\,\propto \,\frac{1}{a^2}$ ) and can be treated theoretically on the basis
of pQCD. On the other hand, the rescattering of gluons
depends on a wider range
of the distances, including the larger ones where
\newline
($r_{\perp}\,>\,\frac{1}{a^2}$). This causes a
 large uncertainty in the pQCD calculations.
 We show that the gluon shadowing generates a damping factor which is
 compatible or bigger than the damping in the quark sector.
\newline
\newline
2.The numerical calculations of the differential cross sections for vector
meson production has been performed using two models for the hadron wave
 function at small transverse distances. The results of the calculations show
that the corrections due to the SC are bigger than the non screened
uncertainties.
There are two sources for these uncertainties: (i) the different
 nonperturbative models for $\Psi^V$ at $r_{\perp} \,\ra\,0$ which, at
 the moment, are not under full theoretical control. And, (ii) the limited
 accuracy of the LLA for the GLAP equation in the available kinematic region.
Basically, those uncertainties can be seen without a detailed calculation
just by changing the scale of $Q^2$ (let say, from $Q^2$ to $Q^2/2$).
However, we think that the exact integration over $z$ in Eq.(9) has a certain
meaning, since the GRV gluon structure function reaches the value of the
anomalous dimension $\gamma = 1/2$ at $x \,\simeq \,10^{-2}$. It means
that at low values of $x$  we can try to calculate the gluon distribution
using the BFKL equation which has a well defined scale in $Q^2$. Therefore,
the difference in
the results
due to the integration over $z$, allows us to estimate the
theoretical uncertainties that originate from our poor knowledge of the
 nonperturbative wave function of the produced vector mesons.
\newline
\newline
3.The numerical calculation of the SC shows that (i) they
should be taken into account in the HERA kinematic region, (ii) their value
is bigger than the uncertainties related to the unknown nonperturbative part
of our calculation, and (iii) DIS vector meson production can be used as a
laboratory for the investigation of SC.
In our opinion, using these reactions to extract the
 nonperturbative wave function of the produced vector mesons
is all too ambiguous, at least with the
present theoretical knowledge of SC.
\newline
\newline
4. The calculation of the SC for the gluon distribution has an intrinsic
uncertainty, related to the large distance contribution. However, we wish
to mention that the anomalous dimension in the GRV gluon distribution
becomes larger than 1 ($ \gamma\,>\,1$) at $x \,<\,10^{-3}$. It means that
at such small $x$, all uncertainties from the large distance behaviour
become rather small, especially at $a^2 \,\leq\,2 GeV^2$ ( which corresponds
 to $Q^2\,\leq\,10 GeV^2$   for the case of $\rho$ production and
$Q^2\,\leq\, 4 GeV^2 $ for $J/\psi$ production). We demonstrated that the poor
accuracy of our calculation in this kinematic region is due to
the fact that the LLA for the GLAP equation is seriously ill in the HERA
region.
Much more work is needed to provide a reliable calculation of the
shadowing corrections for the gluon distribution. We established the fact
that the Mueller (Glauber - like)
formula for the gluon SC cannot provide a reliable
result as  the second iteration of this formula gives
a rather big correction.
We propose to use this formula as an initial condition to solve the new
evolution equations in the region of small $x$ and we will publish a
more detail study of this problem in the future.

5) The relationship between $\frac{\partial F_2^{exp} (Q^2,x)}
{\partial \, ln Q^2}$ and the screened cross section for virtual phtoproduction
of vector mesons has been established. The relation includes all shadowing
corrections and in our oppinion is suitable to extract the information on the
non perturbative light cone $\bar q q$ vector meson wave function. To do this,
we need to know $F_2^{exp} (Q^2,x)$ over a wide region of $x$ and $Q^2$.
With the appearance of the new HERA data
 we hope, that this will be possible in the very near future.

\newpage
\section*{Figure Captions}
\vglue 0.2cm
{\bf Fig.1:}  Vector meson production in DIS without SC.
\newline
\newline
{\bf Fig.2:}  Extra gluon emission for vector meson production
              in DIS without SC.
\newline
\newline
{\bf Fig.3:}  SC for vector meson production in DIS.
\newline
\newline
{\bf Fig.4:}  The time structure of the SC in the inelastic
              cut of the diagram of Fig.3.
\newline
\newline
{\bf Fig.5:}  The behaviour of the damping factor $D^2$ versus
              $x$ = $\kappa_q$ for $J/\psi$ diffractive
              production.
\newline
\newline
{\bf Fig.6:}  ``Fan" diagrams with triple ``ladder" vertices $\g$.
\newline
\newline
{\bf Fig.7:}  The first iteration of the ``fan" diagrams in
              the gluon distribution for vector meson leptoproduction.
\newline
\newline
{\bf Fig.8:}  The x dependence of $\s(\g^* p \ra \rho p)$ with $Q^2$ =
              4, 9, 15, 20 $GeV^2$. a) In a pQCD calculation with no SC.
              b) In a pQCD calculation with SC.
              Solid lines correspond to a CZ wave function input
              and dashed lines to AS.
\newline
\newline
{\bf Fig.9:}  Comparison of screened (solid line) and
              nonscreened (dashed lines) calculated values of
              $\s(\g^* p \ra \rho p)$ for $\sqrt{s}$ = 50, 100, 150 $GeV^2$.
              a) With AS input. b) With CZ input.
\newline
\newline
{\bf Fig.10:}  $\s(\g^* p \ra \rho p)$ as a function of x at $Q^2$ = 8.8 and
               16.9 $GeV^2$ taken from Ref.[33]. Shaded area corresponds to
               a pQCD calculation with ZEUS gluon densities and no SC.
               Solid lines are our AS CZ predictions in a pQCD calculation
               with SC.
\newline
\newline
{\bf Fig.11:}  A mapping of $D^2$ as a function of $\sqrt{s}$ and $Q^2$ for
               $J/\psi$ electroproduction.
\newline
\newline
{\bf Fig.12:}  A compilation of $\s(\g p \ra J/\psi p)$ taken from Ref.[34].
               The solid line is a VDM prediction. The shaded area is the
               prediction of Ryskin[5] with ZEUS gluon densities. The dashed
               lines are our corrections to Ryskin's calculation.
\newline
\newline
{\bf Fig.13:}  The damping factor $D_G^2$ for $J/\psi$ production due to
               first iteration screening in the gluon distribution
               for different values of $a^2$.
\newline
\newline
{\bf Fig.14:}  The ratio
               $R^2 \,= \,D_G^2( second\,\,iteration)
               /D_G^2 (first\,\,iteration)$
               for $J/\psi$ production due to screening in the
               gluon distribution.

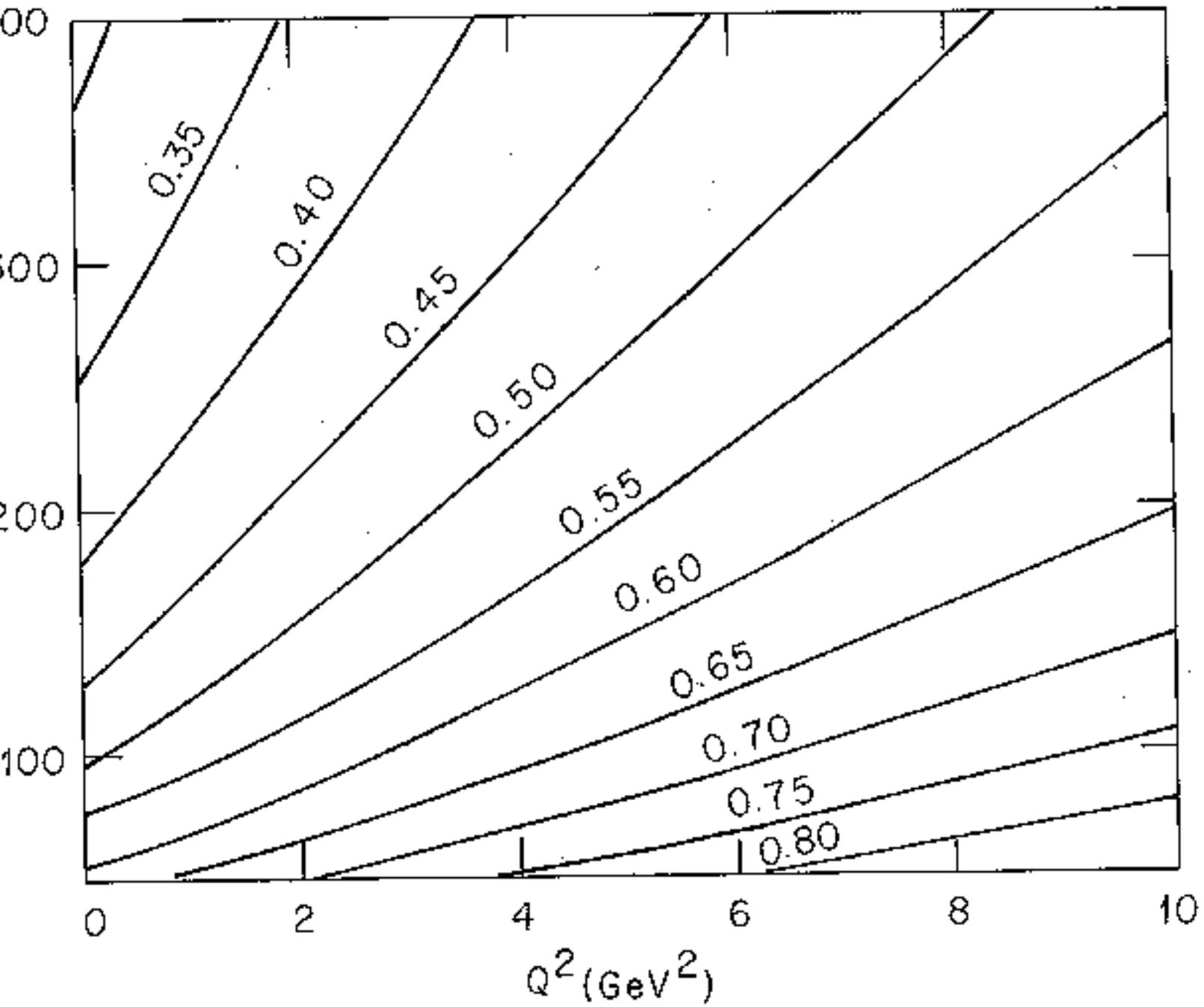
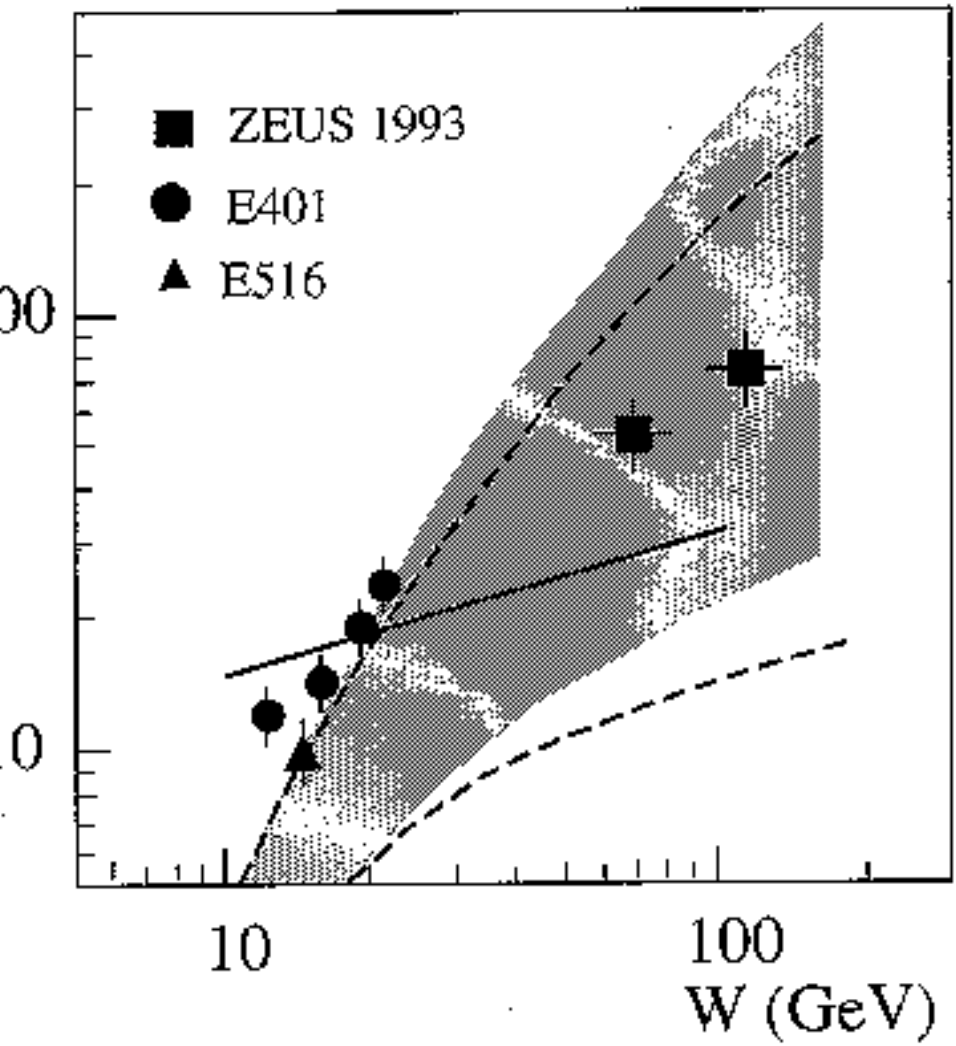
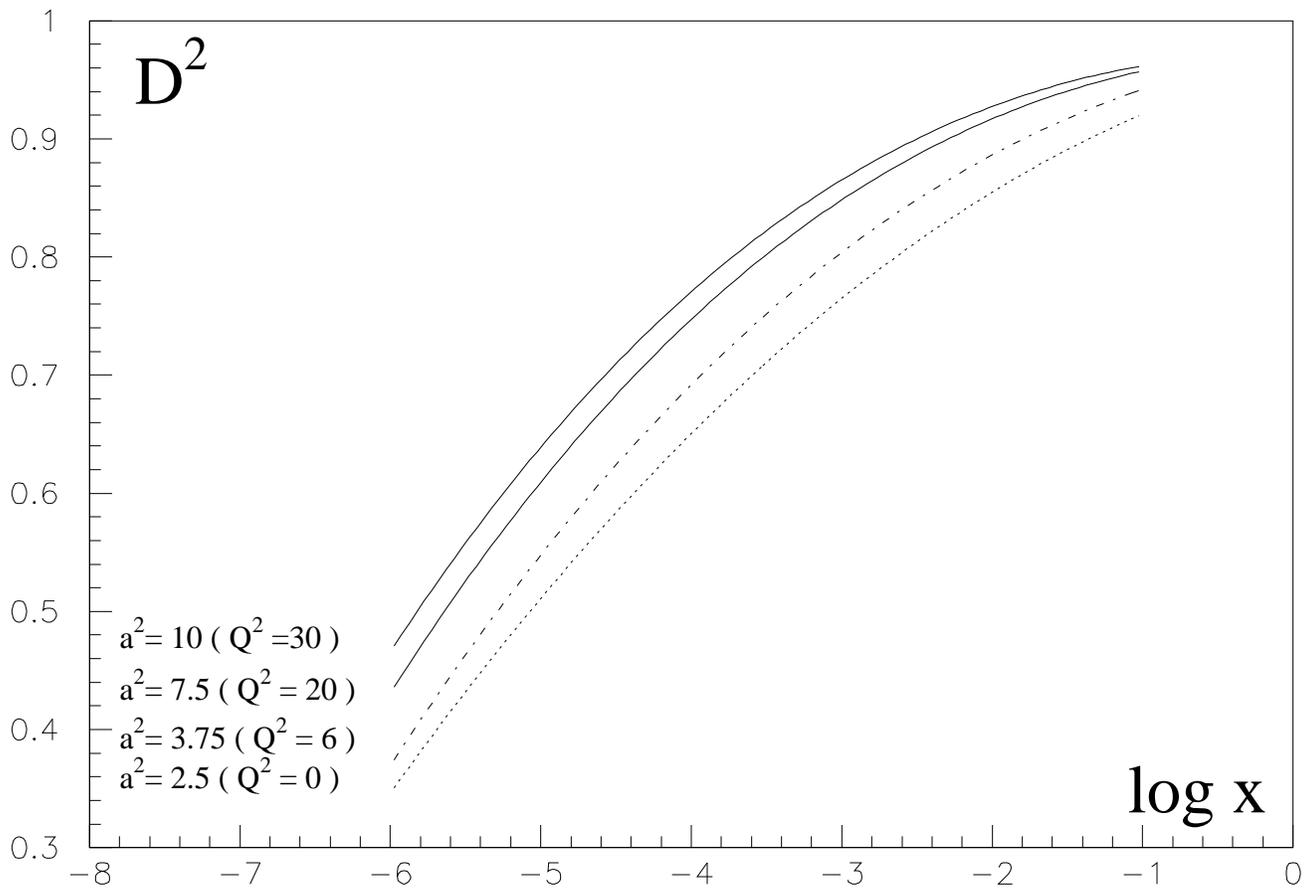
\end{document}